\documentclass{article}

\usepackage{arxiv}

\usepackage[utf8]{inputenc} 
\usepackage[T1]{fontenc}    
\usepackage{hyperref}       
\usepackage{url}            
\usepackage{booktabs}       
\usepackage{amsfonts}       
\usepackage{nicefrac}       
\usepackage{microtype}      
\usepackage{lipsum}

\usepackage[linesnumbered,ruled,vlined]{algorithm2e}
\usepackage{enumitem}
\usepackage{amsfonts}
\usepackage{mathtools}
\usepackage{amsthm}

\newcommand{\tpacas}{TPACAS}
\newtheorem{proposition}{Proposition}
\newtheorem{claim}{Claim}
\newtheorem{theorem}{Theorem}
\newtheorem{definition}{Definition}

\title{A Practical Solution to Yao's Millionaires' Problem and Its Application in Designing Secure Combinatorial Auction}

\author{
  Sankarshan Damle\\
  Machine Learning Lab\\
  IIIT, Hyderabad\\
  \texttt{sankarshan.damle@research.iiit.ac.in} \\
   \And
 Boi Faltings \\
  Artificial Intelligence Laboratory\\
  EPFL, Lausanne\\
  \texttt{boi.faltings@epcl.ch} \\
  \And
  Sujit Gujar\\
  Machine Learning Lab\\
  IIIT, Hyderabad\\
  \texttt{sujit.gujar@iiit.ac.in} \\
}

\begin{document}
\maketitle

\begin{abstract}
The emergence of e-commerce and e-voting platforms has resulted in the rise in the volume of sensitive information over the Internet. This has resulted in an increased demand for secure and private means of information computation. Towards this, the Yao's Millionaires' problem, i.e., to determine the richer among two millionaires' securely, finds an application. In this work, we present a new solution to the Yao's Millionaires' problem namely, Privacy Preserving Comparison (PPC). We show that PPC achieves this comparison in constant time as well as in one execution. PPC uses semi-honest third parties for the comparison who do not learn any information about the values. Further, we show that PPC is collusion-resistance. To demonstrate the significance of PPC, we present a secure, approximate single-minded combinatorial auction, which we call TPACAS, i.e., Truthful, Privacy-preserving Approximate Combinatorial Auction for Single-minded bidders. We show that TPACAS, unlike previous works, preserves the following privacies relevant to an auction: agent privacy, the identities of the losing bidders must not be revealed to any other agent except the auctioneer (AU), bid privacy, the bid values must be hidden from the other agents as well as the AU and bid-topology privacy, the items for which the agents are bidding must be hidden from the other agents as well as the AU. We demonstrate the practicality of TPACAS through simulations. Lastly, we also look at TPACAS' implementation over a publicly distributed ledger, such as the Ethereum blockchain.
\end{abstract}


\section{Introduction}
In the last decade, different \emph{e-commerce} and \emph{e-voting} platforms have grown in popularity. Consequently, the need for privacy of the information exchange within these platforms has become imperative. The consumers (or voters), being strategic agents, prefer the preservation of their \emph{private} information (bids, votes, etc.) as well as their public identities from other (often) competitive agents. This \emph{anonymity} of information also increases participation.

With \emph{blockchain} gaining momentum, e-commerce and e-voting are now being conducted as \emph{smart contracts} over distributed platforms such as \emph{Ethereum}. A smart contract is a computer protocol intended to digitally facilitate, verify, or enforce the negotiation or performance of a contract \cite{wiki:SC}. Since these protocols on the blockchain are on a publicly distributed ledger, they are open to any interested agent, while making the agent's bid/vote as well the execution of its payments publicly verifiable, transparent as well as pseudo-anonymous. However, a consequence of this is that even an agent's private information is publicly available for anyone to see and use. This further necessitates the need for privacy-preserving e-commerce/e-voting protocols, over blockchain. 

At the heart of e-commerce/e-voting protocols is the comparison of two values, either in the form of a bid or a vote. Therefore, in order to build a protocol that preserves each agent's private information, we require a method for comparing these values while preserving their privacy\footnote{The preliminary idea of using secure comparison for designing privacy-preserving auctions has been published in \cite{damle2019truthful}.}. In the literature, this challenge is similar to Yao's Millionaires' problem (Yao \cite{yao1982protocols}) of securely determining the richer between two different parties and has been extensively studied.

    \subsection{Yao's Millionaires' Problem}
Introduced in 1982 by Yao \cite{yao1982protocols}, the Millionaires' problem discusses two agents (millionaires), Alice and Bob, who are interested in knowing the richer among them -- without revealing their true wealth. The first solution to the problem, Yao \cite{yao1982protocols}, is computationally expensive and requires large memory. Thereafter, several protocols with great improvement have been proposed \cite{lin2005efficient,blake2004strong,damgaard2006unconditionally, ioannidis2003efficient}. However, each comparison through these protocols is at best linear in order of the length of the binary representation of these numbers and may also involve multiple rounds of computation. This makes the process computationally expensive for e-commerce/e-voting applications. Further, these protocols require the continuous involvement of agents. Assigning trusted third parties to take part in the protocol on behalf of the agents would reveal the agent's private information to them.

In this paper, we introduce a novel method for comparing two integers $x$ and $y$ (i.e., $x,y\in \mathbf{Z}$) securely, i.e., a solution to the Yao's Millionaires' problem, namely, \emph{Privacy Preserving Comparison} (PPC). In PPC, we assume that there are approved \emph{cryptographic notaries} in the system which act as semi-trusted third parties to assist the \emph{central server} (CS) to determine whether $x\geq y$ or not. We show that neither the CS nor the notaries learn any information regarding the values of $x$ and $y$ during or after the comparison. 

We achieve this secure comparison in constant time, i.e., the complexity of our method is $O(1)$ per comparison. Further, we show that our solution is collusion resistant. We use Pedersen commitment \cite{pedersen1991non} of the values to provide \emph{zero-knowledge proof} for the verifiability of the comparison. We illustrate the utility of our method by using it to build a privacy-preserving protocol for single-minded combinatorial auction, which we call \emph{TPACAS}, i.e., Truthful, Privacy-preserving Approximate Combinatorial Auction with Single-minded bidders.

    \subsection{Secure Combinatorial Auctions}
Auctions are mechanisms which facilitate the buying and selling of goods/items among a group of agents. In general, a combinatorial auction, where the agents can bid for combination(s) of items, yields a higher revenue than selling the items individually. For example, different governments across the globe have been using combinatorial auctions to lease out wireless spectrum \cite{mcmillan1994selling} or allocate airport landing take-off slots to interested agents \cite{rassenti1982combinatorial}. 
In such auctions, the participating entities, which we refer as agents throughout the paper, desire different types of privacy, as the information they submit may expose their business plans to their competitors. For example, the disclosure of an agent's public identity reveals its interest in acquiring the items auctioned. The revelation of an agent's \emph{bidding information} (bid value and the combination of preferred items) to an auctioneer or other participating agents may expose its profits, economic situations and preferences for specific items to its competitors. The competitors may further exploit this information in future auctions. In consequence, an auction protocol should be such that only the winning agents' combination of preferred items is made public while preserving the privacy of the identities and the bidding information of the other agents.

Auction protocols which preserve the privacy of bidding information are called \emph{secure auction} protocols. In this paper, we define these desirable privacies of a secure auction in three types: (i) \emph{Agent privacy}, an agent's participation in an auction must be hidden from all the other agents; (ii) \emph{Bid privacy}, the bid values must be hidden from the other agents as well as the auctioneer; (iii) \emph{Bid-topology privacy}, the items for which the agents are bidding must be hidden from the other agents as well as the auctioneer. 

Furthermore, if the bidding information is hidden from the agents as well as the auctioneer, we need a \emph{trustworthy} implementation of a secure auction. That is, anybody should be able to verify the correctness of the allocations and that the payments are in alignment with the described rules. Besides,  the implementation must preserve all the three types of privacies with high probability. Motivated by these challenges, we focus on the preservation of privacy of all agents' bidding information in an instance of a combinatorial auction.

Typically, the goal in such auctions is to maximize the social welfare, i.e., we should allocate these resources to those who value them most. Strategic agents may misreport their valuations to maximize their profits. Thus, we look for auctions which, through appropriate payment rules, ensure that the agents bid their true valuation. In game theory, such auction protocols (allocation rule along with payment rule) are called \emph{dominant strategy incentive compatible}. In addition to this, auction protocols must also be \emph{individually rational} i.e., protocols wherein the agents have a non-negative payoff.

Combinatorial auctions have an exponential number of possible valuations for each agent and are NP-Complete \cite{rothkopf1998computationally}. Hence, we focus on a \emph{single-minded} case. In this, the agents are interested in a single specific bundle of items and obtain a particular value if they get the whole bundle (or any super-set) and zero otherwise. Even single-minded combinatorial auctions, being NP-Hard \cite{sandholm1999algorithm}, are solved approximately. In particular, Lehmann et al. \cite{lehmann2002truth} propose a strategic proof mechanism for such auctions, which gives $\sqrt{m}$-approximate allocation and payment rule, which we refer to as ICA-SM (Incentive Compatible Approximate auctions for Single-minded bidders). Here, $m$ denotes the number of items being auctioned. In this paper, we propose \tpacas\ (Truthful, Privacy-preserving, an Approximately efficient Combinatorial Auction for Single-minded bidders), which solves a single-minded combinatorial auction, preserving the cryptographic and game theoretic properties mentioned earlier, i.e., \tpacas\ is a trustworthy implementation of ICA-SM.

One can leverage the approach of Micali and Rabin \cite{micali2014cryptography} which uses \emph{homomorphic} property of the commitments or Parkes et al. \cite{parkes2008practical} which also uses \emph{time-lapse cryptography} to achieve winner determination while preserving the privacy of the agents and their bidding information. However, these protocols expose the bidding information to the auctioneer after the bidding phase is over. We overcome these issues by proposing the use of \emph{notaries}. We assume there are approved cryptographic notaries in the system and the auctioneer can appoint them in assisting in the auction. In \tpacas, the auctioneer assigns a signed random $id$ for each agent and a set of randomly chosen notaries. The agents commit their bid values and the size of the bundle in which they are interested similar to \cite{micali2014cryptography}. The challenge remains to sort the bids or to check if two agents have any item in common while keeping the values and bid topology private. Towards this, we use our novel method for secure comparison of two integers, i.e., PPC.

Through this method, we show how to sort as well as compare the bidding information of agents without revealing them, with the help of notaries. The notaries do not learn of any bidding information. We assume that each agent's bundle size is $\geq 2$. Otherwise, bid-topology will get revealed to the auctioneer in our protocol. Note that in our protocol, the notary's role is only to assist the auctioneer in determining winners and their payments when the bidding information is hidden. The notaries will not know the agent identities or their bidding information.

    \subsection{Adversary Model}
    As defined in literature (refer \cite{paverd2014modelling}), and as standard in solutions for Yao's Millionaires' Problem (eg., \cite{blake2004strong,lin2005efficient,grigoriev2017yao,liu2017efficient}),
    in this paper, we assume that \emph{all} agents, i.e., Alice, Bob, auctioneer, notaries etc., are \emph{semi-honest} or \emph{honest-but-curious}. This implies that while these agents can observe and cipher any information, they \emph{do not} deviate from the defined protocol. We use semi-honest and honest-but-curious interchangeably throughout the paper.

\subsection{Contributions}
The following are our contributions: 
    \begin{itemize}
        \item We present a secure, robust and verifiable method, Privacy Preserving Comparison (PPC), for securely comparing two integers (Procedures \ref{procedure:VC} and \ref{ZKP:VC}). We show that the method preserves the privacy of the two integers unless $3$ out of $5$ parties collude.
        \item We propose a cryptographic protocol, \tpacas, that implements a truthful single-minded combinatorial auction (Theorem \ref{theorem:TPACAS}). It is $\sqrt{m}$-approximate and preserves agent privacy of all the losing agents from rest of the agents (Proposition \ref{c1}). It preserves bid privacy with high probability, and the auctioneer will not know any bid value even after the auction is over (Proposition \ref{c2}). It also preserves bid-topology privacy  from the notaries (Proposition \ref{c3}) as well as the auctioneer with high probability (Proposition \ref{c4}).
        \item We believe PPC can be further used to implement other privacy-preserving mechanisms such as other type of auctions, voting etc.
    \end{itemize}
    
    \subsection{Paper Overview}
    The paper is organized as follows: Section \ref{sec:related} discusses the existing results for the Millionaire problem and secure auctions, Section \ref{sec:prelim} describes the relevant cryptographic techniques as well as the auction setting; Section \ref{sec:secure_Comaprison} introduces our novel method for secure comparison of two integers, i.e., PPC, Section \ref{sec:TPACAS} presents the \tpacas\ auction protocol; and Section \ref{sec:Analysis} analyzes it. We conclude and summarize the paper in Section \ref{sec:Conclude}.

\section{\label{sec:related}Related Work}
In this section, we summarize the related literature for (i) Millionaires' Problem; and (ii) Secure Auctions.

    \subsection{Yao's Millionaires' Problem}
    The problem was first introduced by Yao \cite{yao1982protocols} along with its first solution. However, the presented solution is exponential in time and space. After this, most of the solutions Chaum et. al. \cite{chaum1987multiparty}, Beaver and Godwasser \cite{beaver1989multiparty} have focused on using multi-party circuit computations. Grigoriev and Shpilrain \cite{grigoriev2014yao} use various laws of classical physics to present various solutions to the problem. These solutions are irrelevant to an online setting, while also being time consuming for e-commerce/e-voting applications. In this paper, we focus on solutions which can deployed be in an online setting.
    
      Ioannidis and Grama \cite{ioannidis2003efficient} present a two-round protocol which is polynomial while Lin and Tzeng \cite{blake2004strong} and Blake and Kolesnikov \cite{lin2005efficient} provide a single-round solution which is linear in the order of the length of the integers to be compared. For their solutions, \cite{ioannidis2003efficient} uses complex \emph{bitwise operators} while \cite{blake2004strong,lin2005efficient} use \emph{Paillier homomorphic encryptions} and zero-knowledge proof. The computational cost per comparison in \cite{blake2004strong} is $(4b+1)\mbox{(log~}p)+6b$ and in \cite{lin2005efficient} is $5b\mbox{(log~}p)+4b-6$, where $b$ is the bit number and $p$ modulus of the Paillier scheme. Recently, Liu et. al. \cite{liu2017efficient} proposed a single-round solution using Paillier encryption and \emph{vectorization} method. However, the solution is of the order $2(s+2)\mbox{log~}p$, where $s$ is the vector dimension. 
      
      These solutions also require the owners of the integers to do complex operations. Given the number of potential comparisons needed for e-commerce/e-voting applications, continuous involvement of the owners is infeasible. Additionally, one can not simply assign trusted third parties for the operations as that would reveal the owners' private information to them.
      
    \subsection{Secure Auctions}
    VCG mechanisms were proposed by Vickrey \cite{vickrey1961counterspeculation}, Clarke \cite{clarke1971multipart} and Graves \cite{groves1973incentives}. As the allocation problem in a general combinatorial auction is NP-Complete, Lehmann et. al.  \cite{lehmann2002truth} states a strategy proof, approximate greedy mechanism to solve the allocation problem in a restricted setting, without preserving bid privacy. Following the impossibility result on unconditional  privacy Brandt and Sandholm \cite{brandt2008existence}, much of the research has targeted to achieve privacy based on computational hardness of certain problems like discrete-log problem. 

   Micali and Rabin \cite{micali2014cryptography} solves single-item and multi-unit auctions while preserving the privacy of the bids using Pedersen commitment, but reveal the bid information to the auctioneer after the end of the bidding phase, whereas Parkes et. al. \cite{parkes2008practical} uses Paillier encryption and time-lapse cryptography for the same.  \cite{brandt2005efficient} gives a practical, multi-unit auction that does not reveal any private information to a third party, even after the auction closes. Naor et. al.  \cite{naor1999privacy} uses an \emph{auction issuer} while Franklin and Reiter \cite{franklin1996design} uses multiple servers as trusted third parties to solve auctions securely. In both these protocols, the bid-topology is revealed to these third parties. Parkes et. al. \cite{parkes2009cryptographic} uses \emph{clock-proxy} auction to solve a privacy-preserving combinatorial auction, revealing private information to the auctioneer after the end of the clock phase. The protocol is linear in size of the original computational time, from exponential. Suzuki and Yokoo \cite{suzuki2002secure} proposes a privacy-preserving, secure combinatorial auction without revealing any bid information to a third party, using dynamic programming, and \cite{larson2015secure} extends it to add verifiability. The protocol, however, is exponential in size of the number of bids and is thus impractical even for a small number of bids.

\section{\label{sec:prelim}Preliminaries}
In Section \ref{Pre:Sec:CB}, we provide the cryptographic background required for the results; and in Section \ref{Pre:Sec:AD}, we describe the auction setting with the relevant cryptographic and game-theoretic properties.

    \subsection{\label{Pre:Sec:CB}Cryptographic Background}
    For the design of our method for secure comparison of two integers, the following cryptographic techniques are required.
        \subsubsection{Pedersen Commitment Scheme}
            \emph{Commitment functions} are functions that allow one to commit to a chosen value (or chosen statement) while keeping it hidden to others, with the ability to reveal the committed value later.
       
    Let $p$ and $q$ denote large primes such that $q$ divides $p-1$, $G_q$ as the unique subgroup of $\mathbf{Z}^*_p$ of order $q$, and $g$ as a generator of $G_q$. Also, let $g$ and $h=g^a (\mbox{mod~}p)$ be elements of $G_q$ such that $log_gh$ is intractable, where $a \in \mathbf{Z}_q$ is the secret key.
        \begin{definition}
        A Pedersen commitment scheme is the commitment of a message $x \in \mathbf{Z}_q$, with a random help value $r \in \mathbf{Z}_q$, as,
        $$E(x,r)=g^xh^r \:(\textnormal{mod}\:p).$$ 
        \label{PCS}
        \end{definition}
     
        Definition \ref{PCS} follows from \cite{pedersen1991non}. Note that, this commitment scheme is \emph{information theoretically hiding} i.e., given a commitment $E(\cdot)$, every value $x$ is equally likely to be the value committed in $E(\cdot)$; \emph{computationally binding} i.e., an adversary can not find two distinct $x$ and $x'$ which open the same commitment $E(\cdot)$, unless it can solve the discrete-log problem; and is \emph{homomorphic} i.e., given only the public keys ($p,q,g \mbox{~and~}h$) and the commitments of $x_1$ and $x_2$, one can compute the commitment of $x_1 \pm x_2$.
        
           Let $a_i$ denote an agent $i$'s secret key. Thus, every agent $i$'s set of public keys is represented by the set $\mathbb{P}_i=\{p,q,g \mbox{~and~}h_i\}\: \forall i \in A$ .
        \subsubsection{Random Number Representation}
        As standard in the literature (eg., \cite{micali2014cryptography}), we use random number representation of a number $x$, $R(x)$. 
        \begin{definition}
        A random number representation of a number $x$, $R(x)$, is a representation of $x$ as the pair $(u,v)$  where $u,v \in \mathbf{Z}_q$ and $x = (u+v) \:\textnormal{mod}\:q.$
        \label{RNR}
        \end{definition}
        To find $R(x)$ of a number $x$, any agent randomly chooses $u$ and then picks $v=(x-u)\:\textnormal{mod}\:q.$ In this, with only $u$ or $v$, no information about the value of $x$ can be deduced.\\
        
        \noindent\textbf{{Notation}.} In this paper, $E(R(x))$ represents the Pedersen commitment of $x$ as $R(x)=(u,v)$, i.e., $E(R(x))$ denotes the pair of commitments $\big(E(u,r),E(v,r')\big)$.  
        \subsubsection{Zero-knowledge proof}
        In cryptography, Zero-knowledge proof (ZKP) is a method by which an agent, called a \emph{Prover} ($\mathcal{P}$), is able to convince another agent, called a \emph{Verifier} ($\mathcal{V}$), that it knows some information $x$, without revealing $x$ (or any other information related to $x$) \cite{wiki:zkp-property}. Further, $\mathcal{V}$ cannot prove to any other party that $\mathcal{P}$ knows $x$. Informally, ZKP's allows $\mathcal{P}$ to reveal its knowledge of some information, without giving out that information.
        
        In this paper, we model ZKP as an \emph{interaction} (exchange of messages) between $\mathcal{P}$ and $\mathcal{V}$. To this, a ZKP must satisfy the following three properties \cite{wiki:zkp-property}. Here, a honest agent is the one which follows the protocol (proof) correctly.
            \begin{itemize}
                \item \emph{Completeness.} A honest $\mathcal{P}$ will be able to convince a honest $\mathcal{V}$ that the statement is true, if it is true. 
                \item \emph{Soundness.} No dis-honest $\mathcal{P}$ can convince a honest $\mathcal{V}$ that the statement is true, if it is false, with high probability.
                \item \emph{Zero-knowledge.} No $\mathcal{V}$ is able to learn any information regarding the statement, except that it is true, in the case that the statement is true.
            \end{itemize}
        
        \subsubsection{Value Comparison}
        We now look at a method for comparing two values i.e., to find out whether or not $x > y$, when $x,y < q/2$. As shown in \cite{micali2014cryptography}, $x-y\leq q/2 \iff x\geq y$ and $x-y>q/2 \iff x < y$. Therefore, to compare $x$ and $y$ we only need to check whether $x-y\leq q/2$.
        
        \subsubsection{Cryptographic Notaries}
        Similar to \cite{parkes2008practical}, cryptographic notaries are reputable agents, such as law firms, accountants, or firms specializing in providing means of communication of information among agents.
        
        \subsubsection{Secure Information Exchange}
         Similar to \cite{bonneau2014mixcoin}, we define an information exchange as \emph{secure} if it is done over an anonymous and confidential channel. Towards this, \emph{Tor hidden services} \cite{dingledine2004tor} or \emph{SSH connections} \cite{ylonen1996ssh} can be used.
        
    \subsection{\label{Pre:Sec:AD}Auction Design}
    We are considering a situation where an auctioneer ($AU$), the seller itself, is interested in selling $M = \{1,\dots,m\}$ indivisible items and there are $B=\{b_1,\dots,b_{\hat{n}}\}$ $(|B|=\hat{n}\geq 2)$ interested and strategic agents via a combinatorial auction. We assume there exists a set of cryptographic notaries $N$, described in the next section, that can assist $AU$ in determining the winners and their payments. We denote the set consisting of every participating agent in this protocol as $A$ i.e., $A=\left\{\{AU\}\cup B\cup N\right\}$.  

Combinatorial auctions factor in the inter-dependency of the values to an agent with respect to the different combinations possible i.e., each agent has a different preference for different subsets. The valuation function $\vartheta_{b_i}$ describes these preferences $\forall b_i \in B$. In absence of payments, the agent $b_i$ may boast about $\vartheta_{b_i}$. We denote its payment as $\sigma_{b_i}$. Formally, for each possible subset $S \in Q\ \left(Q=2^M\right)$, $\vartheta_{b_i}$ is a real-valued function such that $\vartheta_{b_i}(S)$ is the value an agent $b_i$ obtains if he wins the subset $S$. Also, if $\sigma_{b_i}$ is the price paid by the agent for the subset, then its utility is given by $\psi_{b_i}(\cdot)=\vartheta_{b_i}(S)-\sigma_{b_i}(\cdot)$.

        	\subsubsection{\label{Prop2}Cryptographic Properties in Auction}
    Auction protocols must preserve the privacy of the bidding information from all the agents, including the auctioneer, even after the closing of the bidding phase while providing verifiability of the correctness of the allocation and the payments. In this subsection, we describe these required cryptographic properties of an auction protocol. 
    \begin{itemize}
        	\item \textbf{{Non-repudiation}.} This deals with the inability of an auctioneer or an agent to retract from their actions. 
Auction protocols must be able to commit an agent to its bid as well as prove the exclusion of any bid by the auctioneer.
        \item  \textbf{{Verifiability}.} The public, including the agents, must be shown a conclusive proof of the correctness of the auction protocol. The protocol must enforce correctness; an auctioneer should not be able to present valid proofs for invalid winners or incorrect payments.
          \item   \textbf{{Privacy}.} An auction protocol should hide bidding information of an agent from the other participating agents. After the auction, only the information revealed from the winning agents should be known. The types of privacies relevant for an auction are defined below. For this, let $W$ be the set of winning agents.
            \begin{definition}[Agent Privacy]
                No agent should be able to discover each others identity i.e., for an agent $a \in A$ during the auction and for an agent $a\in A\setminus W$ after the auction, no other agent $b \in A\setminus\{a, AU\}$ should know about $a$'s participation in the auction.    				
            \end{definition}
            \begin{definition}[Bid Privacy] No agent should be able to know any agent's bid valuation i.e., the probability with which an agent $a \in A\setminus\{b_i\}$ can guess agent $b_i$'s bid valuation $\vartheta_{b_i}$ is $\ll 1/\vartheta_{b_i}$.
            \end{definition}
         	\begin{definition}[Bid-Topology Privacy] No agent should be able to know any other agent's bundle of items i.e., the probability with which an agent $a \in A\setminus\{b_i\}$ can guess the item bundle $S_{b_i}$ of an agent $b_i\in B\setminus\{a\}$ during the auction and of an agent $b_i\in B\setminus\{\{a\}\cup W\}$ after the auction is negligible \cite{wiki:NF} in the number of items being auctioned. 
            \end{definition}
    \end{itemize}
Let us say that the allocation of the items is determined by an allocation rule $k(\cdot)$, which takes $\vartheta=(\vartheta_{b_i},\vartheta_{-b_i})$ as the input and outputs who gets which items, where  $\vartheta_{-b_i}$ denotes the set of valuations of agents not including $b_i$. The payment rule is given by $\sigma=(\sigma_{b_1}(\cdot),\sigma_{b_2}(\cdot),\ldots,\sigma_{b_n}(\cdot))$. Thus, an auction is characterized by $(k,\sigma)$, an allocation rule and the payment rule. Given an auction, we need the following game theoretic properties to be satisfied.

        \subsubsection{Game Theoretic Properties in Auction}
                 The valuations of each agent is its private information i.e., hidden from every other agent in the auction. This opens the door for any such agent to lie about their valuations for their benefit. Thus, we look for auctions which incentivize an agent to bid for its true valuation. In mechanism design theory, such truthful auctions are called \emph{ dominant strategy incentive compatible (DSIC)}. Further, an auction is \emph{ex-post individually rational (IR)} if every agent $b_i$ always gets non-negative utility.
             
             \begin{definition}[Dominant Strategy Incentive Compatible]
            An auction $(k,\sigma)$ is DSIC if $\forall \vartheta_{-b_i}$, $\forall \vartheta'_{b_i}$,$\forall b_i \in B$, we have \begin{equation*}
            	\begin{split}
            	\small \vartheta_{b_i}\big(k(\cdot)\big) - \sigma_{b_i}(\vartheta_{-b_i},\vartheta_{b_i}) \geq \vartheta_{b_i}\big(k'(\cdot)\big) - \sigma_{b_i}(\vartheta_{-b_i},\vartheta'_{b_i}) 
            	\end{split}
            \end{equation*} 
              \end{definition}
              where $k(\cdot)=k(\vartheta_{-b_i},\vartheta_{b_i})$ and $k'(\cdot)=k(\vartheta_{-b_i},\vartheta'_{b_i})$.  
            \begin{definition}[Individually Rationality]
            An auction $(k,\sigma)$ is ex-post individually rational if $\forall b_i \in B$, we have $$\vartheta_{b_i}\big(k(\vartheta_{b_1},\dots,\vartheta_{b_{\hat{n}}})\big) - \sigma_{b_i}(\vartheta_{b_1},\dots,\vartheta_{b_{\hat{n}}})\geq 0 \quad \forall \vartheta_{-b_i}.$$
            \end{definition}
    
    As the allocation problem in this setting is \emph{NP-Complete} and because of the difficulty in representing and communicating valuation functions of each agent (since these are exponential in size) in it, we look for much simpler cases of auctions such as the single-minded case.

    \subsubsection{The Single-Minded Case}
    These are auctions wherein agents are interested in a single specific bundle of items, and get a scalar value if they get this whole bundle (or any super-set) and get zero value for any other bundle. Formally, 
    \begin{definition}
    A single-minded valuation function is a function in which there exists a bundle of items $S^*$ and a value $\vartheta^*$ such that $\vartheta(S)=\vartheta^*$, $\forall S \supseteq S^*$ and $\vartheta(S)=0$ for all other $S$. Here, a single-minded bid is the pair $(S^*,\vartheta^*)$.
    \end{definition}
    As the allocation problem in this case is \emph{NP-Hard}, we look at algorithms which can solve this approximately. An algorithm, in an auction setting, is a \emph{$\alpha$-approximation} algorithm if an allocation generated by the algorithm is always less than a factor $\alpha$ times the value of the optimal allocation. We now discuss one such algorithm.
    	\subsubsection{\label{Algo1} An Incentive Compatible approximation Algorithm (ICA-SM)}
        Algorithm \ref{Algorithm:1} describes ICA-SM, which is a \emph{greedy} algorithm that solves the allocation problem for single-minded case with $n$ agents, $m$ items, $\vartheta_{b_i}$ and $S_{b_i}$ as agent $b_i$'s bid valuation and preferred bundle of items, with $W$ as the set of winners approximately. ICA-SM is \emph{computationally efficient}, \emph{incentive compatible} and is $\sqrt{m}$-approximate \cite{lehmann2002truth}. 
           
           \begin{algorithm}[!ht]
\makeatletter
\newcommand{\RemoveAlgoNumber}{\renewcommand{\fnum@algocf}{\AlCapSty{\AlCapFnt\algorithmcfname}}}
\makeatother
        \renewcommand{\algorithmcfname}{Algorithm}
      	 \begin{enumerate}
       \item \emph{Initialization:}
       		\begin{itemize}
       		\item Sort the agents according to the order : $\vartheta_{b_1}^*/\sqrt{|S_{b_1}^*|} \geq \vartheta_{b_2}^*/\sqrt{|S_{b_2}^*|} \geq \dots \geq \vartheta_{b_{\hat{n}}}^*/\sqrt{|S_{b_{\hat{n}}}^*|}$
            \item $W \leftarrow \emptyset$
       		\end{itemize}
       \item \emph{For $i:1\rightarrow \hat{n}$,} if $S_{b_i}^* \cap (\cup_{{b_j} \in W}S_{b_j}^*)=\emptyset$ then $W \leftarrow W \cup \{{b_i}\}$
       \item \emph{Output:} 
       		\begin{itemize}
       		\item \emph{Allocation:} The set of winners is $W$.
            \item \emph{Payments:} $\forall b_i \in W, \sigma_{b_i}=\vartheta_{b_j}^*/\sqrt{|S_{b_j}^*|/|S_{b_i}^*|}$ where $j$ is the smallest index such that $S_{b_i}^* \cap S_{b_j}^* \not= \emptyset$, and for all $k<j$, $b_k\not=b_i$, $S_{b_k}^* \cap S_{b_j}^* = \emptyset$. If no such $j$ exists then $\sigma_{b_i}=0$. 
       		\end{itemize}
       \end{enumerate}
        \caption{\label{Algorithm:1}ICA-SM Algorithm}
    \end{algorithm}
    \normalsize       
       We refer to an auction protocol satisfying all the aforementioned game theoretic and cryptographic properties as a \emph{trustworthy implementation} of an auction, i.e.,
       	\begin{definition}[Trustworthy Implementation]
        \label{trust1}
       	An auction protocol which provides non-repudiation and verifiability, while preserving agent privacy, bid privacy, and bid-topology privacy; along with being dominant strategy incentive compatible and individually rational, is a trustworthy implementation of an auction.
       	\end{definition}

In the next section, we present our method for secure comparison of two integers. In the subsequent subsection, we show the verifiability of the comparison using ZKP.

\section{\label{sec:secure_Comaprison}Privacy Preserving Comparison of Two Integers}

In this section, we first describe a procedure for secure comparison of two integers $x$ and $y$ owned by two agents Alice (say) and Bob (say), respectively. We assume that Alice and Bob have already agreed that $x,y<q/2$. Additionally, we assume that there exists a \textit{central server} (CS) that co-ordinates the comparison. Note that, and as shown later, the CS only aids the comparison and does not learn anything about the values of $x$ and $y$. Procedure \ref{procedure:old_VC} describes the comparison \cite{damle2019truthful}.

Procedure \ref{procedure:old_VC} preserves privacy of the values $x$ and $y$ from CS since CS only knows the values $val_1$ and $val_2$. It is trivial to see that CS shall not be able to find anything about the values of $(u_1,v_1)$ and $(u_2,v_2)$ from these, hence no information about $x$ or $y$ is revealed to it. In addition, every notary only has one component of the other agent's value, which implies that it can not either find out anything about the other agent's value.

 Further, Procedure \ref{procedure:old_VC} is independent of the length of the binary representation of $x$ or $y$ and hence is of constant order ($O(1)$) in computational time. The secure comparison is achieved in one execution of Procedure \ref{procedure:old_VC}.

      \setcounter{algocf}{0}
        \begin{algorithm}
        \DontPrintSemicolon
    \renewcommand{\algorithmcfname}{Procedure}
    CS assigns Alice and Bob their respective pair of \emph{distinct} notaries. Let, $(n^1_{Alice},n^2_{Alice})$ and $(n^1_{Bob},n^2_{Bob})$ be Alice and Bob's pair of assigned notaries, respectively. \newline
     \underline{\textbf{Steps}}\newline
     \begin{enumerate}[label=\roman*]
     \item Alice generates $R(x)=(u_1,v_1)$ and Bob generates $R(y)=(u_2,v_2)$.
     \item 
        $$
        \begin{aligned}
           & \mbox{Alice~} \xrightarrow{u_1}n^1_{Alice}\\
        &    \mbox{Alice~} \xrightarrow{v_1}n^2_{Alice}\\
         &   \mbox{Bob~} \xrightarrow{u_2}n^1_{Bob}\\
          &  \mbox{Bob~} \xrightarrow{v_2}n^2_{Bob}
        \end{aligned}
        $$ All information exchange takes place securely.
    \item $$
        \begin{aligned}
           & n^1_{Bob} \xrightarrow{u_2}n^1_{Alice}\\
        & n^2_{Bob} \xrightarrow{v_2}n^2_{Alice}
        \end{aligned}
        $$
        All information exchange takes place securely.
    \item $$
        \begin{aligned}
           & n^1_{Alice} \xrightarrow{val_1=(u_1-u_2)\mbox{~mod~}q}\mbox{~CS}\\
         & n^2_{Alice} \xrightarrow{val_2=(v_1-v_2)\mbox{~mod~}q}\mbox{~CS}
        \end{aligned}
        $$  All information exchange takes place securely.
     \item  CS then checks the following, \newline
    \emph{if $(val_1+val_2)\mbox{ mod }q=0$ return $\mathbf{``equal"}$} \newline
    \emph{if $ (val_1+val_2)\mbox{ mod }q < q/2$ return $\mathbf{``>"}$} \newline
    \emph{else return $\mathbf{``<"}$}
     \end{enumerate}
    \caption{\label{procedure:old_VC}{Secure Value Comparison of Two Integers}}
	\end{algorithm}

    \noindent\textbf{Discussion.}  By using $x-y$, i.e., $val_1+val_2$, to compare $x$ and $y$ we are able to preserve the privacy of the values $x$ and $y$. However, using Procedure \ref{procedure:old_VC} to design secure mechanisms for applications such as auctions/voting may lead to loss of privacy of the values. For instance, in secure auctions, the bids of the winning agents are opened to determine their payments. This will lead to the disclosure of one value, say $x$, which will consequently disclose $y$ to the CS through the known value $x-y$. Thus, we must also hide the value $x-y$ from the CS while still \emph{preserving} the comparison.

    \subsection{Privacy Preserving Comparison (PPC)}
To overcome the potential loss of the privacy of the values $x$ and $y$ in secure comparison through  Procedure \ref{procedure:old_VC}, we introduce another novel procedure, namely, \emph{Privacy Preserving Comparison} (PPC). Towards this, we assume that Alice and Bob have already agreed that $x,y<\frac{q}{2\cdot D_{{max}}}$. Here, $D_{max}=d^2_{max}$ for $d_{max}=2^{|d_{max}|}$ where $|d_{max}|$ is number of bits required to represent $d_{max}$. In PPC, we require Alice and Bob to privately select an \emph{integer} $d_{Alice},d_{Bob}\in [1,d_{max}]$, respectively. Let, $D=(d_{Alice}\cdot d_{Bob})\mbox{~mod~}q$. 

Before describing PPC, we present the following claim,
    \begin{claim}
    \label{Claim:comparison}
    $ D\cdot(val_1+val_2)\mbox{~mod~}q\leq q/2 \iff x\geq y$ and $D\cdot(val_1+val_2)\mbox{~mod~}q>q/2 \iff x < y$.    
   Here, $R(x)=(u_1,v_1),~R(y)=(u_2,v_2),~val_1=(u_1-u_2)\mbox{~mod~}q,~val_2=(v_1-v_2)\mbox{~mod~}q$ and $D=(d_{Alice}\cdot d_{Bob})\mbox{~mod~}q$ with $x,y<\frac{q}{2\cdot D_{{max}}}$.
    \end{claim}
    \noindent\textbf{Proof.} Observe that,
     $$
    \begin{aligned}
    (x-y)\mbox{~mod~}q&=(u_1+v_1)-(u_2+v_2)\nonumber\\
    &=(u_1-u_2) + (v_1-v_2)\\
    &=(val_1+val_2) \mbox{~mod~}q.
    \nonumber
    \end{aligned}
    $$
 
We know from \cite{micali2014cryptography}: $x-y\leq q/2 \iff x\geq y$ and $x-y>q/2 \iff x < y$. For this, $x,y<q/2$. Now,
    \begin{itemize}[leftmargin=*]
        \item \underline{To show} $D\cdot(val_1+val_2)\mbox{~mod~}q\leq q/2 \iff x\geq y$: Trivially, if $x,y< \frac{q}{2\cdot D}$ and $x \geq y$, we have    
    $$
        (x-y)\mbox{~mod~}q\leq\frac{q}{2\cdot D} \Rightarrow D\cdot (x-y)\mbox{~mod~}q \leq \frac{q}{2}.
    $$
Further, if,
    $$
        D\cdot (x-y)\mbox{~mod~}q \leq \frac{q}{2} \Rightarrow (x-y)\mbox{~mod~}q\leq \frac{q}{2\cdot D} \Rightarrow x \geq y.
    $$
    
    \item \underline{To show} $D\cdot(val_1+val_2)\mbox{~mod~}q>q/2 \iff x < y$: Similarly, if $x,y> \frac{q}{2\cdot D}$ and $x < y$, we have
    $$
        (x-y)\mbox{~mod~}>\frac{q}{2\cdot D} \Rightarrow D\cdot (x-y)\mbox{~mod~}q > \frac{q}{2}.
    $$
    Further, if,
    $$
        D\cdot (x-y)\mbox{~mod~}q > \frac{q}{2} \Rightarrow (x-y)\mbox{~mod~}q> \frac{q}{2\cdot D} \Rightarrow x< y.
    $$
     \end{itemize}
The rest of the claim follows from the fact that as $D_{max}\geq D\Rightarrow\frac{q}{2\cdot D_{{max}}}\leq\frac{q}{2\cdot D}$. \qed\\


    With this claim, we now present our novel method for securely comparing two integers, namely, PPC.
    
    \subsubsection{PPC Procedure}
    
Procedure \ref{procedure:VC} describes the steps taken by Alice and Bob in co-ordination with the CS in PPC. The pair of encryption given at the start and the help values send to the notaries are used for verification of the comparison as shown in Section \ref{Sec:ZKP_VC}. 

        \begin{algorithm}
        \DontPrintSemicolon
    \renewcommand{\algorithmcfname}{Procedure}
    CS assigns Alice and Bob their respective pair of \emph{distinct} notaries. Let, $(n^1_{Alice},n^2_{Alice})$ and $(n^1_{Bob},n^2_{Bob})$ be Alice and Bob's pair of assigned notaries, respectively. \newline
     \underline{\textbf{Steps}}\newline
     \begin{enumerate}[label=\roman*]
     \item Alice generates $R(x)=(u_1,v_1)$ and $E(R(x))$ while Bob generates $R(y)=(u_2,v_2)$ and $E(R(y))$. Then,
    
         $$
        \begin{aligned}
           & \mbox{Alice~} \xrightarrow{E(R(x))}\mbox{~CS}\\
        & \mbox{Bob~} \xrightarrow{E(R(y))}\mbox{~CS}
        \end{aligned}
        $$
     \item 
        $$
        \begin{aligned}
           & \mbox{Alice~} \xrightarrow{u_1,r_1,d_{Alice}}n^1_{Alice}\\
        &    \mbox{Alice~} \xrightarrow{v_1,r_1',d_{Alice}}n^2_{Alice}\\
         &   \mbox{Bob~} \xrightarrow{u_2,r_2,d_{Bob}}n^1_{Bob}\\
          &  \mbox{Bob~} \xrightarrow{v_2,r_2',d_{Bob}}n^2_{Bob}
        \end{aligned}
        $$ All information exchange takes place securely.
    \item $$
        \begin{aligned}
           & n^1_{Alice} \xrightarrow{u_1}n^1_{Bob}\\
        & n^2_{Alice} \xrightarrow{v_1}n^2_{Bob}
        \end{aligned}
        $$
        All information exchange takes place securely.
    \item $$
        \begin{aligned}
           & n^1_{Bob} \xrightarrow{(d_{Bob}\cdot val_1)=(d_{Bod}\cdot (u_1-u_2))\mbox{~mod~}q} n^1_{Alice}\\
         & n^2_{Bob} \xrightarrow{(d_{Bod}\cdot val_2)=(d_{Bod}\cdot (v_1-v_2))\mbox{~mod~}q} n^2_{Alice}
        \end{aligned}
        $$  All information exchange takes place securely.
     \item $$
        \begin{aligned}
           & n^1_{Alice} \xrightarrow{X=D\cdot val_1=(d_{Alice}\cdot d_{Bob}\cdot val_1)\mbox{~mod~}q}\mbox{~CS}\\
         & n^2_{Alice} \xrightarrow{Y=D\cdot val_1=(d_{Alice}\cdot d_{Bob}\cdot val_2)\mbox{~mod~}q}\mbox{~CS}
        \end{aligned}
        $$  All information exchange takes place securely.
     \item  CS then checks the following, \newline
    \emph{if $(X+Y)\mbox{ mod }q=0$ return $\mathbf{``equal"}$} \newline
    \emph{if $ (X+Y)\mbox{ mod }q < q/2$ return $\mathbf{``>"}$} \newline
    \emph{else return $\mathbf{``<"}$}
     \end{enumerate}
    \caption{\label{procedure:VC}{Privacy Preserving Comparison (PPC)}}
	\end{algorithm}

 Procedure \ref{procedure:VC} preserves privacy of the values $x$ and $y$ from CS since CS only knows the values\footnote{For the modular multiplication of $a\cdot b \mbox{~(mod~}q)$, where $q$ is a prime and no information of $a$ is known, all possible values of $b$ are equally likely.} ~$(D\cdot val_1)\mbox{~mod~}q$ and $(D\cdot val_2)\mbox{~mod~}q$. It is trivial to see that CS shall not be able to find anything about the values of $(u_1,v_1)$ and $(u_2,v_2)$ from these, hence no information about $x$ or $y$ is revealed to it. In addition, every notary only has one component of the other agent's (Alice or Bob) value, which implies that it can not either find out anything about the other agent's value. The commitments passed as inputs provide for verifiability of the comparison, as described in the next subsection.

\begin{figure}[h!]
        \centering
        \includegraphics[width=\linewidth]{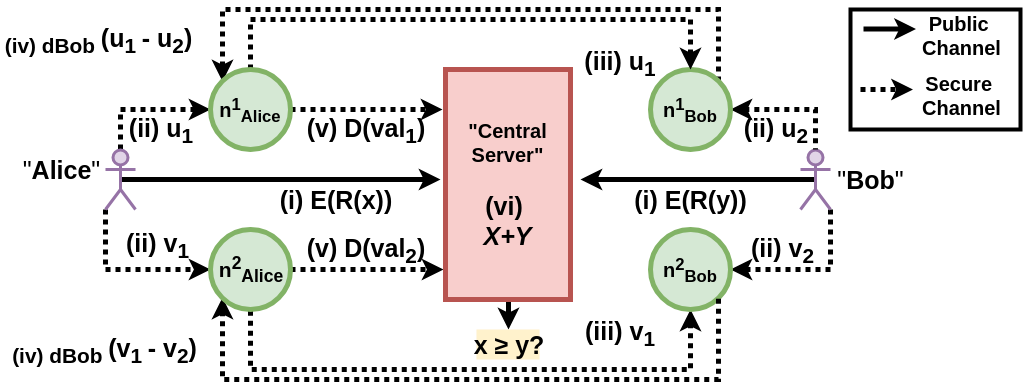}
        \caption{Schematic Representation of the flow of information (i$\rightarrow$ii$\rightarrow$iii$\rightarrow$iv$\rightarrow$v$\rightarrow$vi) during PPC (Procedure \ref{procedure:VC}).}
        \label{fig:sec_comaprison}
    \end{figure}

  
  Note that, PPC is independent of the length of the binary representation of $x$ or $y$ and hence is of constant order ($O(1)$) in computational time. Further, the secure comparison is achieved in one execution of Procedure \ref{procedure:VC}. Figure \ref{fig:sec_comaprison} represents the information flow during Procedure \ref{procedure:VC}, schematically.\\

   \noindent\textbf{Privacy Analysis.} 
   \begin{itemize}
    \item In PPC, by comparing with $X+Y$, we preserve the value of $x-y$. Thus, we make sure that even in the event that one of $x$ or $y$ is revealed, the other value is \emph{not}. Moreover, as $D$ is the product of two random integers owned by Alice and Bob, separately, no one agent can determine the others' value through $X+Y$.
    
    \item Note that, as $d_{max}$ is publicly known, $x-y$ will be an integer in the range $[\frac{X+Y}{D_{max}},X+Y]$. Thus the value of $X+Y$ \emph{bounds} the value of $x-y$. However, the probability of guessing the value of $x-y$, from $X+Y$, can be made negligible by appropriately setting the value for $d_{max}$. For instance, $q$ of the order of $128$ bits and $d_{max}$ of the order $32$ bits, implies that $x-y$ can take values of the order $64$ bits. This results in $2^{64}$ possibilities for $x-y$, i.e., the probability of guessing $x-y$ is negligible.
    
    Moreover, as PPC is independent of the order of $x,y,\mbox{~and~}q,$ one can increase the order of the numbers to further decrease the probability of guessing $x-y$ from $X+Y$ without significantly increasing the computational cost.

     \item  Thus, PPC can be used to design privacy preserving mechanisms like auctions/voting. We illustrate the same with TPACAS (Section \ref{sec:TPACAS}). 
   
    \end{itemize}
   
    \subsubsection{\label{Sec:ZKP_VC}PPC Verification}
Under the assumption that all agents are honest-but-curious, Procedure \ref{procedure:VC} not only preserves the privacy of the values but also ensures the correctness of the comparison. Moreover, the secure comparison is in constant time. We now relax this assumption by assuming Alice and Bob to be strategic agents, i.e., they may misreport the values passed to their assigned notaries in a bid to gain advantage. Thus, in such a setting, we need to \textit{verify} that the values passed during the procedure represent the actual value being compared.

We use ZKP to verify that the values passed to the notaries during Procedure \ref{procedure:VC} are the same as the random representation of $x$ and $y$. We make use of the encryptions\footnote{DSIC mechanisms can be used to ensure that strategic agents report the encryption of their true value (bid/vote etc.) in this step. We illustrate the same with TPACAS (Section \ref{sec:TPACAS}).}, $E(R(x))=\left(E(u_1,r_1),E(v_1,r'_1)\right)$ and $E(R(y))=\left(E(u_2,r_2),E(v_2,r'_2)\right)$, for this. Further, let Alice's public key be denoted by $h_{Alice}$ and Bob's $h_{Bob}$. Procedure \ref{ZKP:VC} describes the interactive ZKP with the CS as the Prover $\mathcal{P}$. For this we also make use of the help values passed by Alice and Bob to their assigned notaries in Step (ii) of Procedure \ref{procedure:VC}. 

We now show that the ZKP described by Procedure \ref{ZKP:VC} satisfies the three properties required for a ZKP, i.e.,
    \begin{itemize}
        \item \emph{Completeness.} It is trivial to see that if Eq. \ref{eqnn::condition} holds, then Eq. \ref{eqn::VC::1} holds. That is, a honest $\mathcal{P}$ will be able convince $\mathcal{V}$ that the comparison was correct.
        \item \emph{Soundness.} If Eq. \ref{eqnn::condition} does not hold, i.e., Alice and/or Bob misreported their values, then there can not be a case where $\mathcal{P}$ can find other values except for $\left(X+Y\right)\mbox{~mod~}q,(H_1)\mbox{~mod~}q\mbox{~and~}\left(H_2\right)\mbox{~mod~}q$ for which Eq \ref{eqn::VC::1} holds, with high probability. This is because Pedersen commitments are computationally binding.\\
        \noindent\textbf{Discussion.} This property also makes the comparison \emph{robust} to any misreporting done by the notaries. Thus, even if we further relax the assumption that the notaries are honest-but-curious, by allowing them to strategically misreport information, Procedure \ref{ZKP:VC} will allow any $\mathcal{V}$ to detect the misreporting. Thus, PPC (Procedures \ref{procedure:VC} and \ref{ZKP:VC}) is \emph{robust} to any misreporting done by the notaries.
        \item \emph{Zero-knowledge.} It is trivial to see that, similar to the argument given for Procedure \ref{procedure:VC}, $\mathcal{V}$ does not gain any knowledge of the committed values or the help values through the values $\left(X+Y\right)\mbox{~mod~}q,H_1\mbox{~mod~}q\mbox{~and~}H_2\mbox{~mod~}q$. Moreover, the value $E(\cdot)^z\mbox{~mod~}p$ does not reveal any information about the value of $z$, at any stage of the procedure, because of the hardness of the discrete-log problem. \qed\\
    \end{itemize}

        \begin{algorithm}[!ht]
        \DontPrintSemicolon
    \renewcommand{\algorithmcfname}{Procedure}
    
         CS has $X$ and $Y$ from Procedure \ref{procedure:VC}.\\
      CS then asks the assigned notaries to compute among them the following,
       $$
        \begin{aligned}
             & n^1_{Alice} \xrightarrow{(d_{Alice}\cdot r_1)\mbox{~mod~}q}n^1_{Bob}\xrightarrow{(D\cdot r_1)\mbox{~mod~}q}\mbox{CS} \\
        & n^2_{Alice} \xrightarrow{(d_{Alice}\cdot r'_1)\mbox{~mod~}q}n^2_{Bob}\xrightarrow{(D\cdot r'_1)\mbox{~mod~}q}\mbox{CS} \\
        & n^1_{Bob} \xrightarrow{(d_{Bob}\cdot r_2)\mbox{~mod~}q}n^1_{Alice}\xrightarrow{(D\cdot r_2)\mbox{~mod~}q}\mbox{CS} \\
         & n^2_{Bob} \xrightarrow{(d_{Bob}\cdot r'_2)\mbox{~mod~}q}n^2_{Alice}\xrightarrow{(D\cdot r'_2)\mbox{~mod~}q}\mbox{CS}
        \end{aligned}
        $$
       Here, $D=(d_{Alice}\cdot d_{Bob})\mbox{~mod~}q$. Further, all information exchange takes place securely.\\

     CS asks $n^1_{Alice}$ to calculate and send the value $E(u_1,r_1)^{d_{Alice}}\mbox{~mod~}p$. CS then asks $n^1_{Bob}$ to calculate and send the value $E(u_1,r_1)^{d_{Alice}\cdot d_{Bob}}\mbox{~mod~}p$. Similarly for commitments $E(u_2,r_2)$, $E(v_1,r'_1)$ and $E(v_2,r'_2)$.\\
     Let,
     $$
     \begin{aligned}
     C&=E(u_1,r_1)^D\cdot E(u_2,r_2)^{-D} \cdot &E(v_1,r'_1)^D \cdot E(v_2,r'_2)^{-D}\mbox{~(mod }p).
     \end{aligned}
     $$ Observe that, 
     \begin{align}  
    C &= E(u_1,r_1)^D\cdot E(u_2,r_2)^{-D}\cdot E(v_1,r'_1)^D\cdot\nonumber E(v_2,r'_2)^{-D}\nonumber \\
        &= \left(g^{u_1} h_{Alice}^{r_1}\right)^D\cdot\left( g^{u_2} h_{Bob}^{r_2}\right)^{-D}\cdot \left(g^{v_1} h_{Alice}^{r'_1}\right)^D\cdot\nonumber\left( g^{v_2} h_{Bob}^{r'_2}\right)^{-D}\nonumber\\
        &=g^{D\cdot (u_1-u_2+v_1-v_2)}\cdot h_{Alice}^{D\cdot (r_1+r'_1)}\cdot  h_{Bob}^{-D\cdot(r_2+r'_2)}\nonumber\\
  C&=g^{(X+Y)\textnormal{ mod }q}\cdot h_{Alice}^{H_1\textnormal{~mod~}q}\cdot h_{Bob}^{H_2\textnormal{ mod }q}\mbox{~(mod }p)\label{eqn::VC::1}
 \end{align}
 
        $\mathcal{V}$ accepts that 
        \begin{equation}
        \begin{aligned}
        \label{eqnn::condition}
        \begin{split}
           (X+Y)\mbox{ mod }q&=\left(D\cdot (x-y)\right)\mbox{ mod }q\\
           H_1\mbox{ mod }q&=\left(D\cdot (r_1+r'_1)\right)\mbox{ mod }q\\
           H_2\mbox{ mod }q&=-\left(D\cdot (r_2+r'_2)\right)\mbox{ mod }q\\
         \end{split}\quad\Bigg \rbrace
        \end{aligned}
        \end{equation} only if Eq. \ref{eqn::VC::1} holds.
        
    \caption{\label{ZKP:VC}{ZKP for PPC}}
	\end{algorithm}


    \subsection{Collusion in PPC} Procedure \ref{procedure:VC} and Procedure \ref{ZKP:VC} provide a secure and verifiable way for comparing two integers. The comparison requires $5$ semi-honest third parties -- $4$ notaries and $1$ CS. While Procedure \ref{ZKP:VC} ensures that the method is \emph{robust} to misreporting of any value, collusion among the agents may result in loss of privacy of the values.
    
    For instance, if both the assigned notaries to Alice/Bob collude, the privacy of the integer owned by Alice/Bob is lost. However, this form of collusion is difficult in a real-world setting since the notaries will not be aware of each others existence in the comparison. This is because, from Procedure \ref{procedure:VC} or Figure \ref{fig:sec_comaprison}, there is no line of communication between two notaries of the type $n_j^i, \forall i\in \{1,2\}$ where $j$ is either Alice or Bob. 
    
    However, if \emph{any one} of the two assigned notaries to Alice as well as Bob collude with CS, i.e., $3$ out of $5$ parties, then the privacy of both the integers is lost. This follows similar to other third party secure protocols like \cite{rivest2014practical,suzuki2002secure,shamir1979share}. Thus, PPC is collusion resistant unless the CS is part of the collusion.
    
    \subsection{PPC Illustration}
    We now illustrate the verifiable comparison of two integers in PPC (Procedures \ref{procedure:VC} and \ref{ZKP:VC}) with an example. Let, the values to be compared be $7$ owned by Alice, and $6$ owned by Bob with $d_{max}=5$. Further, let $p=1187,~q=593,~g=3,~a_{Alice}=2,~a_{Bob}=3,~h_{Alice}=9\mbox{~and~}h_{Bob}=27$. Let, $d_{Alice}=2\mbox{~and~}d_{Bob}=3$. We have $7,6<\frac{593}{2\cdot5^2}$. 
    
    With this, $7$ can be represented as the pair $(350,250)$ with $E(R(7))=\left((350,11),(250,4)\right)$ and $6$ can be represented as the pair $(300,299)$ with $E(R(6))=\left((300,12),(299,15)\right)$. The following steps describe the verifiable comparison in PPC.
        \begin{itemize}
            \item Alice and Bob send $E(R(7))$ and $E(R(6))$ to CS.
            \item Alice sends the pair of values $(350,11\cdot2,2)$ to $n^1_{Alice}$ and the pair of values $(250,4\cdot2,2)$ to $n^2_{Alice}$. Similarly, Bob sends $(300,12\cdot3,3)$ to $n^1_{Bob}$ and $(299,15\cdot3,3)$ to $n^2_{Bob}$.
            
            \noindent\textbf{{Comparison}.}
            \item $n^1_{Alice}$ sends $350$ to $n^1_{Bob}$ and $n^2_{Alice}$ sends $250$ to $n^2_{Bob}$.
            \item $n^1_{Bob}$ sends $3\cdot 50$ to $n^1_{Alice}$ and $n^2_{Bob}$ sends $3\cdot 544$ to $n^2_{Alice}$.
            \item $n^1_{Alice}$ calculates $D\cdot val_1=2\cdot 3\cdot 50$ or $D\cdot val_1\mbox{~mod~}q=300$ and $n^2_{Alice}$ calculates $D\cdot val_2=2\cdot 3\cdot 544$ or $D\cdot val_2\mbox{~mod~}q=299$. The notaries send their respective values to the CS.
            \item CS calculates $D\cdot (val_1+val_2)\mbox{~mod~}q= X+Y\mbox{~mod~}q=300+299=6<q/2$. Hence, CS returns the result $7>6$.
            
            \noindent\textbf{{Verification}.}
            \item CS shows $\mathcal{V}$ that 
                $$
                \begin{aligned}
             C &= \left(3^{350}\cdot3^{22}\right)^6\cdot \left(3^{300}\cdot3^{36}\right)^{-6}\cdot\\&\left(3^{250}\cdot3^{8}\right)^6\cdot \left(3^{299}\cdot3^{45}\right)^{-6} \\
             &= 3^{453}\cdot 3^{356}\cdot 3^{362}\cdot 3^{308} \mbox{~(mod~}1187)\\
             & =  410\cdot 33\cdot 317\cdot 682 \mbox{~(mod~}1187)=899\\
                \end{aligned}
                $$
            \item CS asks the assigned the notaries to send the values  $D\cdot (r_1),~D\cdot (r'_1),~D\cdot (r_2)
            \mbox{~and~} D\cdot (r'_2)$ as described in Procedure \ref{ZKP:VC}. CS already has $X$ and $Y$.
            \item We have $D\cdot (r_1)=66,~D\cdot (r'_1)=24,~D\cdot (r_2)=72
            \mbox{~and~} D\cdot (r'_2)=90$.
            This information exchange takes place securely.
            \item CS computes $H_1=D(r_1+r_1')=(66+24)\mbox{~mod~}593=90$ and $H_2=-D(r_2+r_2')=-(72+90)\mbox{~mod~}593=431$.
            \item CS shows the following,
            $$
            \begin{aligned}
            g^{X+Y}\cdot h_{Alice}^{H_1}\cdot h_{Bob}^{H_2} &=3^{6}\cdot 9^{90}\cdot 27^{431}\mbox{~(mod~}1187) \\
            &=729\cdot 592\cdot 181\mbox{~(mod~}1187)\\
            & = 899=C\\
            \end{aligned}
            $$
             \item Hence, $\mathcal{V}$ is convinced that the comparison was for the same values as the ones committed by Alice and Bob.\\
        \end{itemize}

    We now use PPC introduced for secure comparison of two integers to present a novel, secure combinatorial auction for the single-minded case that preserves the privacy of each agent's bidding information even after the bidding phase is over, namely, \tpacas.

\section{\label{sec:TPACAS}\tpacas\ Auction Protocol}
In \tpacas, $A$ is set of agents wherein $AU$ is the seller itself, and all arithmetic operations (except the payments) are modulo $p$ for the commitments and modulo $q$ for the values to be committed as well as the help values. Further, $AU$ acts as the CS. We assume that $AU$ and the set of notaries are honest-but-curious, while the bidders are also strategic as described in Section \ref{Sec:ZKP_VC}. Before describing the secure auction protocol, we define the following with respect to TPACAS:
    \begin{itemize}
        \item \textbf{{Item Bundle}.} In \tpacas, an agent $b_i$ submits its \emph{item bundle} $\mathbb{S}_{b_i}$, consisting of commitments of its preferred items \emph{at least} once as well as \emph{different} commitments of some (or all) of their preferred items randomly such that $|\mathbb{S}_{b_i}|=m,\ \forall b_i \in B$. Formally,
        	\begin{definition}[Item Bundle]
        	\label{IB1}
            An agent $b_i$'s item bundle is defined as $\mathbb{S}_{b_i}=\left\{C\cup D\right\}$ where  $C=\{E(R(j))\:|\:\forall j \in S_{b_i}\}$ and $D=\{E(R(k))\:|\:\forall k \in S'_{b_i}\}$,
        	\end{definition}
        where $S'_{b_i}$ is the set of non-distinct items randomly chosen from $S_{b_i}$ such that $|C| + |D|=m$.
        \item \textbf{{Secure Bulletin Board}.} Secure Bulletin Boards (SBB) consists of publicly known websites which are controlled by $AU$. All data published is time stamped and cannot be erased. $AU$ uses the SBB to publish all public information about the auction, including the initial auction announcement as well as (committed) information that have been submitted and proofs that can be used to verify all publicly available information about the outcome. The content of the SBB is viewable to all participating agents -- and all are assured that they are viewing the same content. For example, the SBB can be a smart contract over blockchain, since all the values submitted on a smart contract will be on a publicly distributed ledger such as on the Ethereum blockchain.
 \end{itemize}

\begin{figure}
	\centering
    \includegraphics[width=\columnwidth]{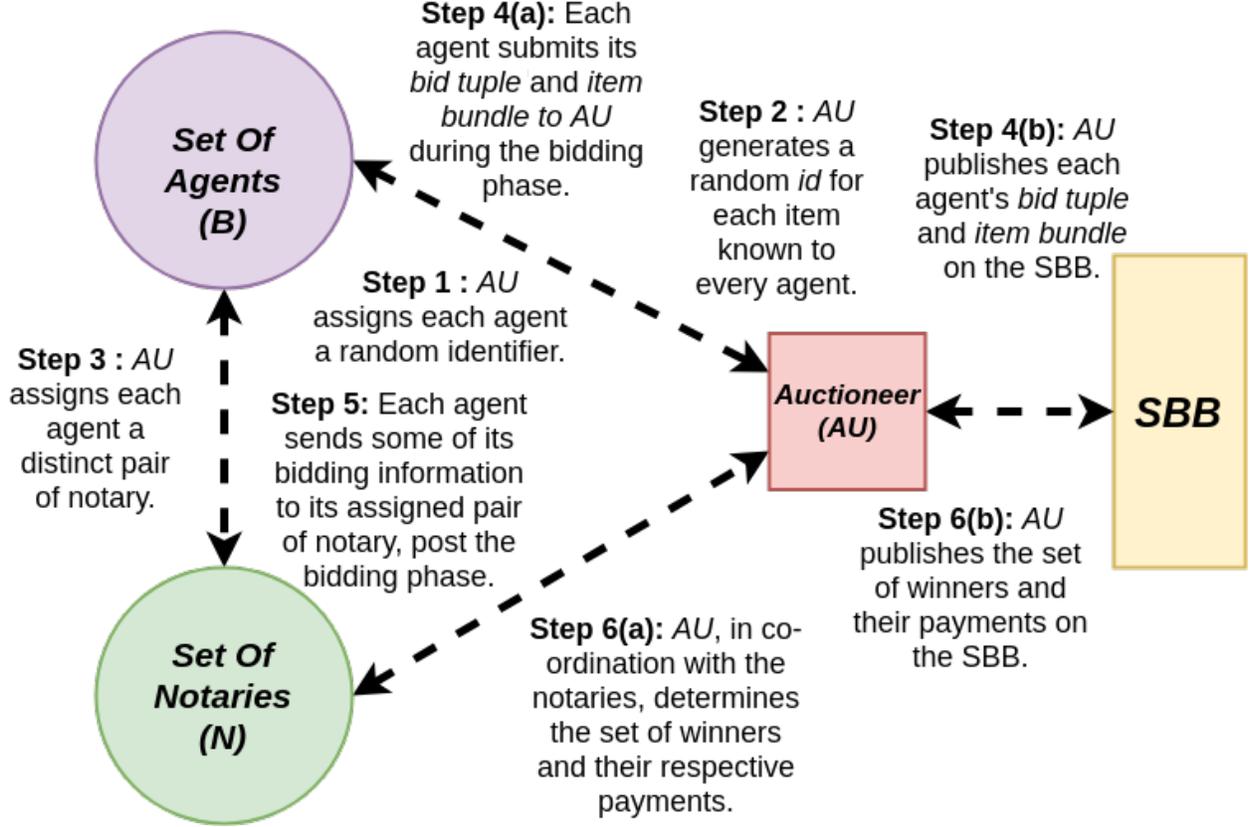}
    \caption{\label{fig:tpacas_protocol_figure}Schematic representation of the information flow during the execution of \tpacas.}
\end{figure}

 Protocol \ref{Protocol:1} illustrates the \tpacas\ auction protocol presented. Note that, while we require $d_{id_{b_i}}\in [1,d_{max}],\forall id_{b_i}$, we do not require any such bound\footnote{This step ensures that no information about the items being compared is revealed (Section \ref{Sec:WD}).} on $d'_{id_{b_i}}, \forall id_{b_i}$.
 Figure \ref{fig:tpacas_protocol_figure} summarizes  the information flow among the participating agents during the execution of the protocol, schematically.

 \setcounter{algocf}{0}
    \begin{algorithm}[!ht]
 \DontPrintSemicolon
   
\makeatletter
\newcommand{\RemoveAlgoNumber}{\renewcommand{\fnum@algocf}{\AlCapSty{\AlCapFnt\algorithmcfname}}}
\makeatother
        \renewcommand{\algorithmcfname}{Protocol}
        $AU$ sets up the auction by announcing $p,q,g,\mbox{~and~}d_{max}$ (or $D_{max}$) as well as the items being auctioned.\\
        At the start of the auction, every agent $a \in A \setminus \{AU\}$ gives its public \emph{id}'s to $AU$ upon which $AU$ assigns to every agent $a$ a secret identifier $id_a$, securely. These are known only to $AU$ and \emph{not} to any other agent.\\
        $AU$ generates a random $id$ for each item which is known to every  agent but \emph{not} to notaries. The agents commit these \emph{ids} instead of directly committing their preferred set of items. The $ids$ can be greater than $q/2$, as to compare items we only need to check if they are equal. \\
        $AU$ assigns a pair of notaries $(n_{id_{b_i}}^1, n_{id_{b_i}}^2) \in N$ to every agent $id_{b_i} \in B$ randomly.\newline
        \noindent\textbf{\underline{Bidding Phase}:} \\
       (a) $\forall \: id_{b_i} \in B$, submits its \emph{bid tuple}, i.e., $\mathbf{BT}_{id_{b_i}}=\big\langle E(\vartheta_{id_{b_i}}),E(|S_{id_{b_i}}|),E\big(R(\vartheta_{id_{b_i}}/\sqrt{|S_{id_{b_i}}|})\big)$ to $AU$. Each agent's $w_{id_{b_i}}=\vartheta_{id_{b_i}}/\sqrt{|S_{id_{b_i}}|}$ must be less than $\frac{q}{2\cdot D_{max}}$. Every $id_{b_i}$ also submits its item bundle $\mathbb{S}_{id_{b_i}}$. (b) $AU$ publishes the bid tuple and the item bundle on the SBB for non-repudiation. \newline
        \noindent\textbf{\underline{Post-Bidding Phase}:} \\
        Each agent $id_{b_i} \in B$ sends  the random representations of the value $w_{id_{b_i}}$ as well as the random representation of all the commitments in $\mathbb{S}_{id_{b_i}}$, along with the  help values of their commitments and their private integers to their assigned notaries, i.e., $id_{b_i}$ sends $\left(u,r,d_{id_{b_i}},d'_{id_{b_i}}\right)$ to $n_{id_{b_i}}^1$ and $\left(v,r',d_{id_{b_i}},d'_{id_{b_i}}\right)$ to $n_{id_{b_i}}^2$ for the value $w_{id_{b_i}}$ as well as its item bundle. All information exchange takes place securely. \\
        (a) $AU$ determines -- in co-ordination with the assigned notaries -- the set of the winning agents $W$, consisting of each winner's identifier, and calculates payments as described in ICA-SM, and (b) publishes it on the SBB.
       \caption{\label{Protocol:1}\tpacas\ Auction}  
    \end{algorithm}
    
    \subsection{\label{BIPhase}Bid Initialization}
     $AU$ sorts the bids based on the values $w_{id_{b_i}}$ $\forall id_{b_i}$, using any comparison based sorting with the comparison done through Procedure \ref{procedure:VC} with the private values $d_{id_{b_i}},\forall {id_{b_i}}$.
     
      For the winner and payment determination phase, the highest agent's identifier is denoted as $id_{b_1}$, the second highest agent's as $id_{b_2}$, and so on. Let $I$ consist of the set of identifiers $\{id_{b_1},\dots,id_{b_{\hat{n}}}\}$, $\mathbb{S}$ as the set of preferred item bundles of every agent $\{\mathbb{S}_{id_{b_1}},\dots,\mathbb{S}_{id_{b_{\hat{n}}}}\}$ and $W$ as the set of winners initialized to $\emptyset$.
    
    \subsection{\label{Sec:WD}Winner Determination}
       $AU$ carries out winner determination, as described in Algorithm \ref{Algorithm:1}, in co-ordination with notaries. In this, the highest agent is automatically selected, and its \emph{identifier} is added to $W$. To determine the other winners, $AU$ compares every pair of item, $\forall \: id_{b_j} \in I \setminus \{id_{b_1}\}$ with every $id_{b_k}$ currently in $W$, using Procedure \ref{procedure:VC} with the private values $d'_{id_{b_i}},\forall {id_{b_i}}$.
       
        If $AU$ does not find any identical pair of items for an agent $id_{b_j}$ for every  $id_{b_k}$ currently in $W$ i.e., $\mathbb{S}_{id_{b_j}} \cap (\cup_{k \in W}\ \mathbb{S}_{id_{b_{k}}})=\emptyset$, it adds $id_{b_j}$ to $W$. Otherwise, it discards that agent and continues with the next highest agent. \\

    \noindent\textbf{Discussion.} As the set of items $M$ is finite, i.e., there are only $\binom{m}{2}$ distinct combinations possible, $AU$ can deterministically get the items, $x$ and $y$, being compared from the value $val_1+val_2 $, i.e., $x-y$. By using PPC and imposing no restriction\footnote{We do not need to restrict the value of $d_{max}$ for item comparison, as we are only interested in knowing whether the items are the same, i.e., whether $x-y=0$ or not. This ensures that $X+Y$ does not leak the bound for $x-y$.} on the values $d'_{id_{b_i}},\forall {id_{b_i}}$, however, if $x\not= y$, i.e., $X+Y\not =0$, all possible $\binom{m}{2}$ combinations will be equally likely.

    \subsection{Payment Determination}
    The payments for every winner $id_{b_i} \in W$ are as described in Algorithm \ref{Algorithm:1}. $AU$ can find out an agent $id_{b_j}$, $\forall \: id_{b_i} \in W$, where $j$ is the smallest index such that $\mathbb{S}_{id_{b_i}} \cap \mathbb{S}_{id_{b_j}} \not= \emptyset$, and an agent $id_{b_k}$ for $k<j$, $id_{b_k}\not=id_{b_i}$ such that $\mathbb{S}_{id_{b_k}} \cap \mathbb{S}_{id_{b_j}} = \emptyset$, similar to the procedure to the winner determination described in Section \ref{Sec:WD}. If such $id_{b_j}$ and $id_{b_k}$ exists, then $AU$ asks the assigned notary $n^1_{id_{b_j}}$ of $id_{b_j}$ to calculate the payment $\sigma_{id_{b_i}}=\vartheta_{id_{b_j}}/\sqrt{|S_{id_{b_j}}|/|S_{id_{b_i}}|}$. The agent $id_{b_j}$ opens its commitment $E(R(w_{id_{b_j}}))$ for $n^1_{id_{b_j}}$, securely. $AU$ asks $id_{b_i}$ to open its commitment for $E(|S_{id_{b_i}}|)$, and sends the value to $n^1_{id_{b_j}}$, which calculates $\sigma_{id_{b_i}}$ and sends it to $AU$. If no such  $id_{b_j}$ or $id_{b_k}$ exist, then $\sigma_{id_{b_i}}=0$.
    
\section{\label{sec:Analysis}TPACAS Analysis}
It is trivial to see that \tpacas\ preserves non-repudiation. We now look at verifiability and the nature of the privacy guarantees as provided by \tpacas. In this section, we denote the identifier $id_{b_i}\in B$ as $b_i$ for simplicity of notation. 

    \subsection{\label{TPACAS:VER}Verifiability}
    A prover $\mathcal{P}$ ($AU$) proves to a verifier $\mathcal{V}$  the correctness of the order $w_{{b_1}} \geq \dots \geq w_{{b_{\hat{n}}}}$ and the correctness of the comparisons for $\mathbb{S}_{{b_i}} \cap \mathbb{S}_{{b_j}} = \emptyset$, for each ${b_i},{b_j} \in B$.
    
    As all value \emph{as well as} item comparisons in \tpacas, are done using PPC (Procedure \ref{procedure:VC}), the ZKP for the comparisons follows the same as described in Procedure \ref{ZKP:VC}. \qed\\
    
   \noindent\textbf{Discussion.}	Since Pedersen commitments are computationally binding, $\mathcal{V}$ is sure of the comparison without the need to be shown multiple proofs for different commitments of the same values, for value \emph{as well as} item comparisons. This significantly reduces the computational time as compared to other secure auction protocols such as \cite{parkes2008practical}. We further analyze this in Section \ref{Sec:Simulation}.
    
    \subsection{Privacy Analysis}
            \tpacas\ provides the following privacy guarantees:
    	\begin{proposition}
        \label{c1}
    	\tpacas\ preserves agent privacy.
    	\end{proposition}
    	    \noindent\textbf{Proof.} In \tpacas, first all the agents communicate with $AU$ over a secure channel indicating their interest in the auction. $AU$ assigns a random $id$ to all the agents which are used through-out the execution of \tpacas\ for communication. Thus, there is no step in the protocol from which the agent identities will be leaked unless $AU$ leaks the identities.
        
        In case the SBB is a smart contract, agent privacy is preserved by leveraging the fact that public addresses of participating agents are pseudo-anonymous i.e., these addresses can not be linked to the real-world identity of the agents with high probability. \qed

        \begin{proposition}
        \label{c2}
        \tpacas\ preserves each agent's bid privacy.
        \end{proposition}
\noindent\textbf{Proof.} Each agent is required to submit only the Pedersen commitments of their bids, which are information-theoretically hiding. Hence, it is impossible for any malicious agent to open a Pedersen commitment to reveal a message $x \in \mathbf{Z}_q$, other than through a brute force attack. Let $|q|$ be the number of bits required to represent the prime $q$, then the probability that an agent can open a commitment this way is of the order $1/2^{|q|}$.
 
 Typically, $64$ bits are enough to represent each agent's bid valuation while $|q|$ is chosen to be $1024$ bits. Thus, we have $1/2^{|\frac{q}{2\cdot D_{max}}|}\ll 1/\vartheta_{b_i}~\forall b_i$.
 \qed
        \begin{proposition}
        \label{c3}
        \tpacas\ preserves bid and bid-topology privacy from the notaries.
        \end{proposition}
\noindent\textbf{Proof.} Since each individual notary will only know one value (either $u$ or $v$) of its assigned agent ${b_i}$'s $w_{{b_i}}$ or $\mathbb{S}_{{b_i}}$ and assuming they do not collude among each other or with $AU$, \tpacas\ preserves the bid and bid-topology privacy from the notaries. \qed


\begin{proposition}
        \label{c4}
        In \tpacas, the probability with which $AU$ can know at least one item in agent $b_j$'s bid-topology is $1/s_{b_i}$. The probability with which $AU$ can know the complete bid-topology of an agent $b_j$ is, 
        \begin{equation}
           \label{eq1}
        P_{b_j}(s_{b_i}) = \frac{1}{2^{m}-2^{m-s_{b_i}}}
        \end{equation}
         $\forall b_j \in B\setminus W$, such that $b_i\in W$ is that agent for which $S_{b_j} \cap S_{b_i}\not=\emptyset$ in Step 2 of Algorithm 1, $s_{b_i}=|S_{b_i}|$ and $m$ is the number of items.
        \end{proposition}
             \noindent\textbf{Proof.}
        $AU$, through the bidding topology of the winners and its knowledge about which agents have at least one item in common, can infer some information about the bid-topology of an agent $b_j \in B\setminus W$.
        
        With this, $AU$ can know that $b_j$'s bid-topology consists of at least one item belonging to $b_i$'s preferred bundle of items, $S_{b_i}$. The probability with which $AU$ can figure that item is $1/s_{b_i}$.
        
        Further, with this information, $AU$ can eliminate all possible subsets of $M$ which do not consist of any item in $S_{b_i}$. Then the probability with which \tpacas\ leaks each agent $b_j$'s bid-topology to $AU$ is given by Eq. \ref{eq1}, as all the remaining subsets i.e., $2^{m}-2^{m-s_{b_i}}$ are equally likely. \qed

    From Eq. \ref{eq1}, \tpacas\ preserves bid-topology privacy with high probability when $s_{b_i}\geq 2, \forall b_i \in W$. For the analysis of the result, observe that Eq. \ref{eq1} can be written as,
    \begin{equation*}
    \begin{split}
    P_{b_j}(s_{b_i}) 
    &= \frac{1}{2^{m}-2^{m-s_{b_i}}}\\
    &=\frac{1}{2^m}\left(\frac{1}{1-1/2^{s_{b_i}}}\right)\\
    &=\frac{2^{s_{b_i}}}{2^{s_{b_i}}-1}\left(\frac{1}{2^m}\right).
    \end{split}
    \end{equation*}
    Thus, the increase in the probability with which $AU$ can determine the complete bid-topology of an agent with respect to randomly guessing the complete bid-topology is by a \emph{constant factor}, i.e., by $\frac{2^{s_{b_i}}}{2^{s_{b_i}}-1}$. Assuming that each agent's bundle size is $\geq 2$, the worst case follows when $s_{b_i}=2$. The probability that $AU$ can know the complete bid-topology of an agent $b_j$ in this case is,
    $$P_{b_j}(s_{b_i})=\frac{4}{3}\left(\frac{1}{2^m}\right),$$
which is an increase by a factor $\frac{4}{3}$ or an increase by $33.33\%$ of $O(\frac{1}{2^m})$ which is negligible in $m$.

        The probability result follows from the fact that at no point during the auction or post-auction and $\forall b_j \in B\setminus W$, the cardinality of the preferred bundle of items of an agent $b_j$ i.e., $s_{b_j}$, is revealed to $AU$ in \tpacas. Note that, Eq. \ref{eq1} does not hold for an auction protocol that leaks the cardinality of $S_{b_j}$ of an agent $b_j$. For instance, if $AU$ knew that for an agent $b_j$, $s_{b_j}=m$, the probability with which agent $b_j$'s bid-topology is leaked to $AU$ would be $1$.

\subsubsection{Collusion} In \tpacas, the participating agents may collude amongst one other, in an effort to change the outcome of the auction or know the bidding information of other agents. Since every assigned notary receives any bidding information of its assigned agent after the end of the bidding phase; and because \tpacas\ satisfies non-repudiation, the agents in collusion with it cannot alter any of their submitted information. Further, due to verifiability and non-repudiate property of \tpacas, any set of colluding agents cannot strategically change or ask an agent to withdraw any submitted information. Thus, collusion in \tpacas\ cannot affect the outcome of the auction. Such a collusion may, however, reveal bidding information of agent(s) depending on the type and the number of agents which are colluding. For instance, if both the assigned notaries of an agent ${b_i}$ i.e., 3 out 5 parties in Procedure \ref{procedure:VC}, colludes with $AU$, the bid-topology of agent ${b_i}$ will be revealed to $AU$. Further, as $D$ will be different for each comparison, if at least one notary from every pair ($n$ out of $2n$) colludes with $AU$, $AU$ will know the bid-topology of every agent. This follows similar to other third-party secure protocols such as \cite{suzuki2002secure} and \cite{rivest2014practical}.

Note that the protocol is resilient to collusion among the notaries, as there is no line of communication between two notaries of the type $n_{b_j}^i, \forall i\in \{1,2\}$ for each $b_j\in B$.

        \begin{theorem} 
        \tpacas\ is trustworthy implementation of ICA-SM.
        \label{theorem:TPACAS}
        \end{theorem}
\noindent\textbf{Proof.} Propositions \ref{c1}, \ref{c2}, \ref{c3} and \ref{c4} show that \tpacas\ preserves agent, bid and bid-topology privacy with high probability (the probablity of guessing improves only by $O(\frac{1}{2^m})$) and is non-repudiate and verifiable. Since the protocol also solves the winner and payment determination problem through ICA-SM, it is DSIC and ex-post IR. Thus, the theorem follows from Definition \ref{trust1}. \qed
    
    \subsection{Implementation}
    We now look at some implementation aspects of \tpacas. For implementation, to avoid any floating-point number, $AU$ can announce at the start of the auction, that the value $w_{{b_i}}$ for every agent ${b_i}$, will have \emph{$x$-precision} i.e., each value $w_{{b_i}}$ will be significant up to $x$ decimal places.

  \begin{table}
   \renewcommand{\arraystretch}{1.1}
   \centering
	\begin{tabular}{ |c|c|c|c|c| }    
\hline
$\hat{n}$ & $m$ & Upper Bound & Mean & Mean Time  \\
~&~&$(\sqrt{m})$&$\bigg(\frac{Opt \:Welfare}{Approx \:Welfare}\bigg)$ & Taken $(min)$\\
\hline 
25 & 9 & 3 & 1.11905993576 & 2.1826\\
25 & 12 & 3.4641 & 1.1313692063 & 5.21355 \\
25 & 15 & 3.8729 & 1.05711039103 & 11.103467\\
\hline
100 & 9 & 3 & - & 11.59642 \\
100 & 12 & 3.4641 & - & 19.72178 \\
100 & 15 & 3.8729 & - & 54.084380 \\
\hline
\end{tabular}
\caption{\tpacas\ approximation bound for 25 random auction instances for each.}
\label{tab:ratio_table}
\end{table}

    \subsubsection{\label{Sec:Simulation}Simulation Analysis}
    For the implementation, all auction instances were generated as CATS file using SATS command line tool \cite{weiss2017sats}. The optimal social welfare was calculated by solving the winner determination problem for the general single-minded case through FRODO 2.0 \cite{leaute2009frodo}. For this, the generated file was parsed through the inbuilt FRODO 2.0 parser to convert CATS to XCSP.  This parsed file can be solved using any of the optimal algorithms (such as DPOP, P-DPOP etc.) provided in FRODO 2.0 (through GUI or command line).
    
    The primes $p\mbox{~and~}q$ were of size 1024 bits. Table \ref{tab:ratio_table} shows the results. Note that, for large $\hat{n}$ it is difficult to calculate the optimal welfare, as the problem is NP-Hard. The results show the practicality of the trade-off in \tpacas\ i.e., the protocol is not losing significant revenue for a substantial decrease in computational time.

The mean time taken for \tpacas\ in Table \ref{tab:ratio_table} includes  the verification of every value and item comparison done throughout the execution of \tpacas. The verification of these comparisons (Procedure \ref{ZKP:VC}) is the only computationally expensive operation in \tpacas. However the time consumed for verification of the value as well as item comparisons, is significantly less as compared to other secure auction protocols such as \cite{parkes2008practical}. 

For comparison, a 100 bid \emph{single item} auction (i.e., $\hat{n}=100$ and $m=1$) takes approximately $2.51$ hours in \cite{parkes2008practical} (see \cite[Table 2]{parkes2008practical}), while a $100$ bid \emph{single minded} combinatorial auction (i.e., $\hat{n}=100$) even with $m=15$, only takes approximately $0.91$ hours in TPACAS.

    \subsubsection{Implementation over Blockchain}
    
    \begin{figure}[ht!]
	\centering
    \includegraphics[width=\columnwidth]{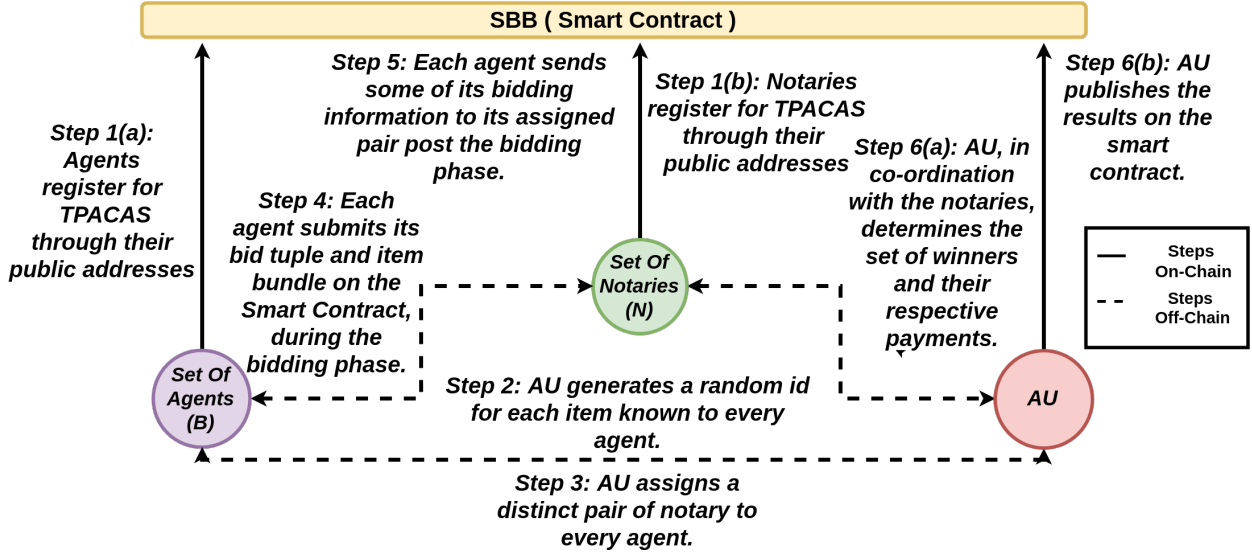}
    \caption{\label{fig:tpacas_protocol}Schematic representation of the flow of information in \tpacas, when the SBB is a smart contract. 
    }
\end{figure}
    Towards this, the SBB will be implemented as a smart contract. The agents (including the notaries) register themselves for the auction through their public addresses, which acts as their random $id$'s. $AU$ generates a random $id$ for the items. $AU$ sends the $id$'s to every agent securely and off-chain. Further, $AU$ assigns each agent a pair of distinct notaries off-chain. The agents submit their bid tuples and their item bundle on the smart contract. The agents also submit some of their bidding information, as described in the \tpacas\ Protocol, off-chain to their assigned notaries.

For the winner and payment determination, $AU$, through communication with the notaries determines the set of winners off-chain. Figure \ref{fig:tpacas_protocol} illustrates the information flow during execution of the TPACAS protocol over blockchain, schematically. In Figure \ref{fig:tpacas_protocol}, the solid arrows represent steps conducted on-chain, i.e., Step 1(a), Step 1(b), Step 4 and Step 6(b). The dashed arrows represent steps conducted off-chain, i.e., Step 2, Step 3, Step 5 and Step 6(a).\\

    	\noindent\textbf{{Notary Fees}.} We suggest to pay the notaries by implementing the SBB as a smart contract over blockchain, which shall act as an \emph{incentive} for them to not alter with any information, since that would be detected and they would not be paid.

\section{\label{sec:Conclude}Conclusion}
We have presented a practical, robust and verifiable solution to the Yao's Millionaires' Problem, namely, PPC (Procedures \ref{procedure:VC} and \ref{ZKP:VC}). PPC uses semi-honest third parties to securely compare two integers, who do not learn any information. PPC achieves this comparison in constant time as well as in one execution of Procedure \ref{procedure:VC}. Further, we show that PPC is collusion resistant.

To demonstrate the significance of PPC, we use it to design a secure combinatorial auction auction protocol, namely, \tpacas. \tpacas\ is practical, strategic proof, verifiable and preserves bidding information with high probability. \tpacas\ preserves an agent's bid valuation as well as bid-topology at any time during the auction and post-auction, even to the auctioneer, unlike previous secure combinatorial auctions. The bid-topology is preserved with high probability when every agent's bundle size is $\geq 2$, which is a fair assumption in practice for combinatorial auctions. Also, we illustrate how to execute \tpacas\ over a publicly distributed ledger such as blockchain. 

We believe that PPC can be further used to implement other privacy-preserving mechanisms including different types of auctions, voting etc.

\bibliographystyle{unsrt}  
\bibliography{references}  

\end{document}